\documentclass[reprint,twocolumn,aps,prc,a4paper,superscriptaddress,preprintnumbers,amsmath,amssymb]{revtex4}
\usepackage{color}
\usepackage{graphicx}
\usepackage{CJK}
\usepackage{longtable,booktabs}
\usepackage{ltablex} 
\begin{document} 
\title{Finding the best basis states for the variation after projection nuclear wave functions}

\author{Xiao Lu}
\email{xlusdu@163.com}
\affiliation{China Institute of Atomic Energy, P.O. Box 275 (10), Beijing 102413, China}
\author{ Zhan-Jiang Lian}
\affiliation{China Institute of Atomic Energy, P.O. Box 275 (10), Beijing 102413, China}
\author{Xue-Wei Li}
\affiliation{China Institute of Atomic Energy, P.O. Box 275 (10), Beijing 102413, China}
\author{ Zao-Chun Gao}
\email{zcgao@ciae.ac.cn}
\affiliation{China Institute of Atomic Energy, P.O. Box 275 (10), Beijing 102413, China}
\author{ Yong-Shou Chen}
\affiliation{China Institute of Atomic Energy, P.O. Box 275 (10), Beijing 102413, China}

\begin{abstract}
 The variation after projection (VAP) method is expected to be an efficient way of getting the optimized nuclear wave functions, so that they can be as close as possible to the exact shell model ones. However, we found there are two additional problems that may seriously affect the convergence of the VAP iteration. The first problem is, if a randomly selected projected basis state does not mix with a VAP wave function in the VAP calculation, then it is likely that this basis state will never mix with the VAP wave function even after the VAP iteration converges, which means such selected projected basis state is useless. The other problem is the poor orthonormality among the projected basis states that seriously affect the accuracy of the calculated VAP wave function. In the present work, solutions for these two problems are proposed and some examples are presented to test the validity. It turns out that, with the present solutions, the most important projected basis states can be reliably obtained and the fully optimized VAP wave functions can be accurately and efficiently calculated. ~ 
\end{abstract}

 

\maketitle
\section{Introduction}
Nuclei are complicated quantum many-body systems. According to the quantum mechanics, the wave functions of low-lying nuclear states should be obtained by solving Schr\"{o}dinger's equation. Practically, this is done by performing the full shell model calculation in a given model space. However, the configuration space can be easily huge in a large model space, which makes the full shell model calculation almost impossible. This difficulty pushes theorists to develop approximated shell model methods, so that the obtained nuclear wave functions are expected to be as close as possible to the exact shell model ones. So far various approximated shell model methods, such as the shell model truncation \cite{ST94}, stochastic quantum Monte Carlo approaches and their extrapolations \cite{MC97,MC01,MC12,MCE10,MCE21},
projected configuration interaction \cite{PCI09}, and the class of variation after projection (VAP) methods \cite{VAMPIR04,Shimizu21,VAP15,VAP17,VAP18,VAP22}, have been well developed.

Among those approximated shell model methods, the VAP is an important one, which is also believed to have a good shell model approximation in calculating low-lying states of nuclei \cite{VAMPIR04,Shimizu21,VAP15,VAP17,VAP18,VAP22}. Such approximation can be continuously improved by adding more and more projected basis states into that calculated states. Certainly, the added projected states should be important so that the calculated states can be significantly improved. However, the problem is, if an added projected state is randomly generated and does not mix with a calculated state at the beginning of the VAP iteration, then it is likely that this added projected state will never mix with the state even after the VAP iteration converges. This means such added projected basis state is useless and must be abandoned. In this sense, a useful new projected state should mix with a calculated nuclear state at the beginning of the VAP iteration. Therefore, an effective way of how to get a new useful projected states before performing the VAP iteration is crucial in the VAP calculation.

This problem may be solved by previous methods, such as the quantum Monte Carlo diagonalization (QMCD) \cite{MC01}. In the QMCD, one selects the best new basis state from the stochastically generated candidates. To ensure the effectiveness of the new candidate, shell model hamiltonian is diagonalized in a subspace spanned by all previously selected basis states and the candidate basis. One can then check
the contribution of this candidate basis for lowering the energy eigenvalue being calculated. If such contribution is large enough, then this candidate basis will be added to the group of basis states. Otherwise, one needs to check the importance of the next basis candidate. After that, only important basis states were selected and a good approximation is expected. However, in such basis selection, one might sometimes check a large amount of useless candidates before a new important basis is identified. Completely different from the QMCD, here, we propose a new reliable and efficient algorithm, in which a randomly generated useless projected state can be varied so that it may become a new important basis state for the VAP calculation.

The second problem in the VAP calculation is the poor orthonormality among the projected basis states, which seriously affects the accuracy of the calculated VAP wave functions and consequently damages the stability of the VAP iteration. This problem has been considered in our previous work \cite{VAP22}, where the projected states are imposed by two constraints to prevent the appearance of redundant states throughout the VAP iteration. In the present work, we replace those two constraints with a new one, which is expected to be more efficient in keeping the projected basis states in good condition during VAP calculation.

This paper is organized as follows. Section \ref{BS} provides a solution of how to generate the useful projected basis states. Section \ref{ON} discusses the problem of orthonormality among the projected basis states.  Section \ref{56ni} shows an example of the present VAP calculations. A brief summary and outlook are presented in Section \ref{sum}.

\section{The important projected basis states}\label{BS}

Let's first address the problem of the important projected basis states in the VAP calculation. We start from the simplified VAP wave function taken from our previous work \cite{VAP22},
\begin{eqnarray}\label{wf_n}
|\Psi^{(n)}_{J\pi M\alpha}(K)\rangle=\sum_{i=1}^n f^{J\pi\alpha}_{i}P^{J\pi}_{MK}|\Phi_i\rangle,
\end{eqnarray}
where $|\Phi_i\rangle$ is a Slater determinant (SD) so that the particle number projection can be omitted here. $n$ is the number of include $|\Phi_i\rangle$ SDs. $P_{M K}^{J\pi}$ stands for the product of the angular momentum projection operator, $P_{M K}^{J}$, and the parity projection operator, $P^\pi$. The $K$ number can be randomly chosen in the range of $|K|\leq J$. $\alpha$ is used to differ the states with the same $J$, $\pi$ and $M$.  The $f^{J\pi\alpha}_{i}$ coefficients and the corresponding energy, $E^{(n)}_{J\pi\alpha}$, are determined by the following Hill-Wheeler equation,
\begin{eqnarray}\label{hw}
\sum_{i'=1}^{n}(H^{J\pi}_{ii'}-E^{(n)}_{J\pi\alpha}N^{J\pi}_{ii'})f_{i^{\prime}}^{J\pi\alpha}=0,
\end{eqnarray}
where $H_{i i'}^{J\pi}=\langle \Phi_i|\hat H P^{J\pi}_{KK}|\Phi_{i'}\rangle$ and $N_{i i'}^{J\pi}=\langle \Phi_i|P^{J\pi}_{KK}|\Phi_{i'}\rangle$. The normalization condition is imposed on the $f^{J\pi\alpha}_{i}$ coefficients so that $\langle\Psi^{(n)}_{J\pi M\alpha}(K)|\Psi^{(n)}_{J\pi M\alpha}(K)\rangle=1$.

\begin{figure}[ht!]
 \centering
 \includegraphics[width=8cm]{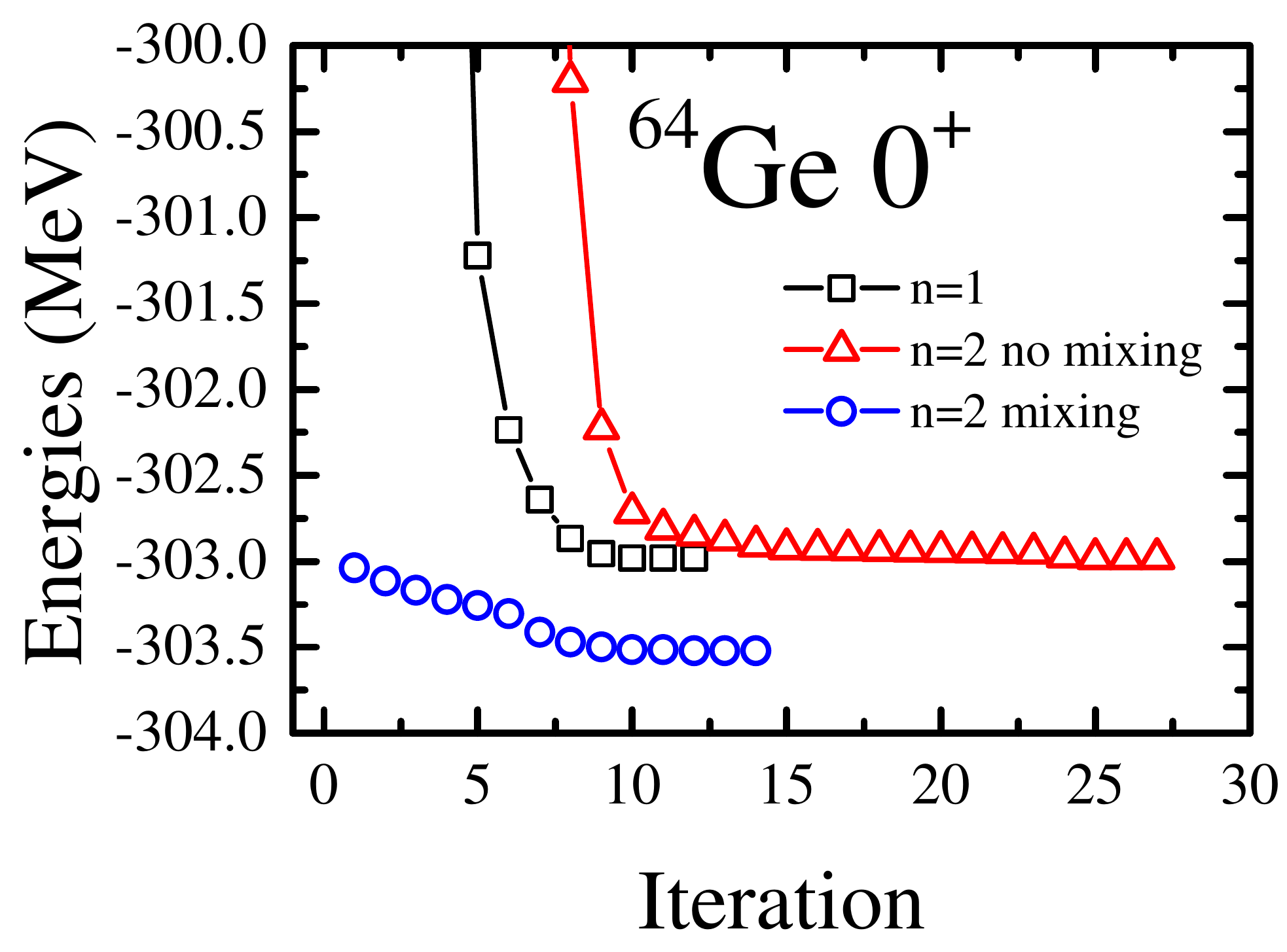}
 \caption{\label{Ge64}(Color online) The VAP iterations with n=1 (black square), n=2 without mixing (red triangle) and n=2 with mixing (blue circle) for $J^\pi=0^+$ in $^{64}$Ge.
}
\end{figure}

Naturally, one can directly take the trial wave function in Eq. (1) to perform the VAP iteration with the initial $|\Phi_i\rangle$ SDs randomly generated. As a simple example, we use the trial wave function in Eq. (\ref{wf_n}) with $K = 0$ to perform the VAP calculations for the ground $0^+$ state in $^{64}$Ge. The GXPF1A interaction \cite{gxpf1a} in the $fp$ model space is taken. Thus the parity projection can be omitted in this example. The results are shown in Fig. \ref{Ge64}. It is seen that in the simplest case of $n=1$, the VAP iteration converges quite fast and the converged energy is $-302.983$MeV. However, if one takes $n=2$ with random initial SDs, it is quite possible that the final converged VAP energy is also $-302.983$MeV, which is the same as that with $n=1$. This strongly implies that one of the two projected SDs does not mix with the converged VAP wave function at all.

Theoretically, if there is a new state, $|\phi\rangle$, that does not mix with a known wave function $|\psi\rangle$, then one can easily prove the following identity,
\begin{eqnarray}\label{mix}
\langle \phi|(\hat H-E)|\psi\rangle=0,
\end{eqnarray}
where $E=\langle\psi|\hat H|\psi\rangle$. Actually, $|\phi\rangle$ is usually not orthogonal to $|\psi\rangle$. So, $|\phi\rangle$ needs to be orthogonalized by performing the Gram-Schmidt orthogonalization, and the orthogonalized one can be written as
\begin{eqnarray}\label{gram}
|\phi'\rangle=|\phi\rangle-\langle\psi|\phi\rangle|\psi\rangle,
\end{eqnarray}
so that $\langle\psi|\phi'\rangle=0$.
If there is no mixing between $|\phi\rangle$ and $|\psi\rangle$, then one should have
\begin{eqnarray}\label{cp}
\langle \phi'|\hat H|\psi\rangle=0.
\end{eqnarray}
By substituting $|\phi'\rangle$ into Eq. (\ref{cp}), one can immediately get Eq. (\ref{mix}).

Therefore, one can simply introduce a real and nonnegative quantity that can be used to indicate the coupling strength between the normalized $|\phi\rangle$ and $|\psi\rangle$ states,
\begin{eqnarray}
c=|\langle \phi|(\hat H-E)|\psi\rangle|^2.
\end{eqnarray}
Certainly, if $c>0$, then $|\phi\rangle$ definitely mixes with $|\psi\rangle$. Thus one may need to consider how to find a $|\phi\rangle$ so that the $c$ value can be large enough.

In the present work, we take the $|\Psi^{(n)}_{J\pi M\alpha}(K)\rangle$ wave function in Eq. (\ref{wf_n}) as $|\psi\rangle$ and the new candidate projected state $\frac{P^{J\pi}_{MK}|\Phi\rangle}{\sqrt{\langle\Phi|P^{J\pi}_{KK}|\Phi\rangle}}$ as $|\phi\rangle$. Therefore, one can calculate the following quantity
\begin{eqnarray}\label{ca}
c_\alpha=\frac{|\langle \Phi|P^{J\pi}_{KM}(\hat H-E^{(n)}_{J\pi\alpha})|\Psi^{(n)}_{J\pi K\alpha}(K)\rangle|^2}{\langle \Phi|P^{J\pi}_{KK}|\Phi\rangle},
\end{eqnarray}
for each calculated $|\Psi^{(n)}_{J\pi M\alpha}(K)\rangle$ state.

  More generally, when one calculates $m$ lowest states simultaneously using the algorithm in Ref. \cite{VAP18}, then the candidate projected state is useful as long as one of the $c_\alpha$ quantities is large enough. Hence one can define a global $C$ quantity as,
\begin{eqnarray}\label{cg}
C=\sum_{\alpha=1}^m c_\alpha.
\end{eqnarray}
It is easy to understand that if the $C$ value is large enough, then the candidate projected state should be important for the calculated states.

Now, let us try to solve the problem of the example in Fig. \ref{Ge64}. Based on the converged VAP wave function with $n=1$, whose energy is $-302.983$MeV, we then randomly generate a second projected state $P^{J\pi}_{MK}|\Phi\rangle$ and combine it with the projected basis in this $n=1$ VAP wave function to form a $n=2$ wave function, which explicit form can be written as
\begin{eqnarray}\label{wf2}
|\Psi^{(2)}_{J\pi M1}(K)\rangle=f^{J\pi 1}_{1}P^{J\pi}_{MK}|\Phi_1\rangle+f^{J\pi 1}_{2}P^{J\pi}_{MK}|\Phi\rangle.
\end{eqnarray}
This $|\Psi^{(2)}_{J\pi M1}(K)\rangle$ wave function will be further optimized fully by simultaneously varying both $|\Phi_1\rangle$ and $|\Phi\rangle$ SDs, so that improved energy minimum lower than $-302.983$MeV is expected.

However, if a candidate projected state is randomly generated in a huge configuration space, it is likely the $C$ value is extremely tiny. Indeed, in the present case, the calculated $C$ value between the random $P^{J\pi}_{MK}|\Phi\rangle$ and the converged $|\Psi^{(1)}_{J\pi M1}(K)\rangle$ is as tiny as $10^{-10}$, which is almost equals to zero. This leads to the fact that the second term in Eq. (\ref{wf2}) can be neglected, which means the $f^{J\pi 1}_{2}$ coefficient can be very tiny. Consequently, the gradient of the energy for the wave function in Eq. (\ref{wf2}) should be small enough to terminate the VAP iteration at the very beginning. Therefore, the VAP wave function can not be further improved.

 This forces us to find out an algorithm to vary the candidate projected state so that the $C$ quantity can be maximized. To make such maximization more efficient, it is necessary to calculate the gradient and Hessian of $C$ with respect to the variational parameters of $|\Phi\rangle$. Fortunately, the matrix elements required by such gradient and Hessian are available in the present VAP calculations. This makes it very convenient in the maximization of the $C$ quantity. Since such maximization of the $C$ quantity  is equivalent to the minimization of the $-C$, in the practical calculation, we prefer to take the latter so that our present algorithm of the minimization in the VAP calculation can be directly adopted.

\begin{figure}[ht!]
 \centering
 \includegraphics[width=8cm]{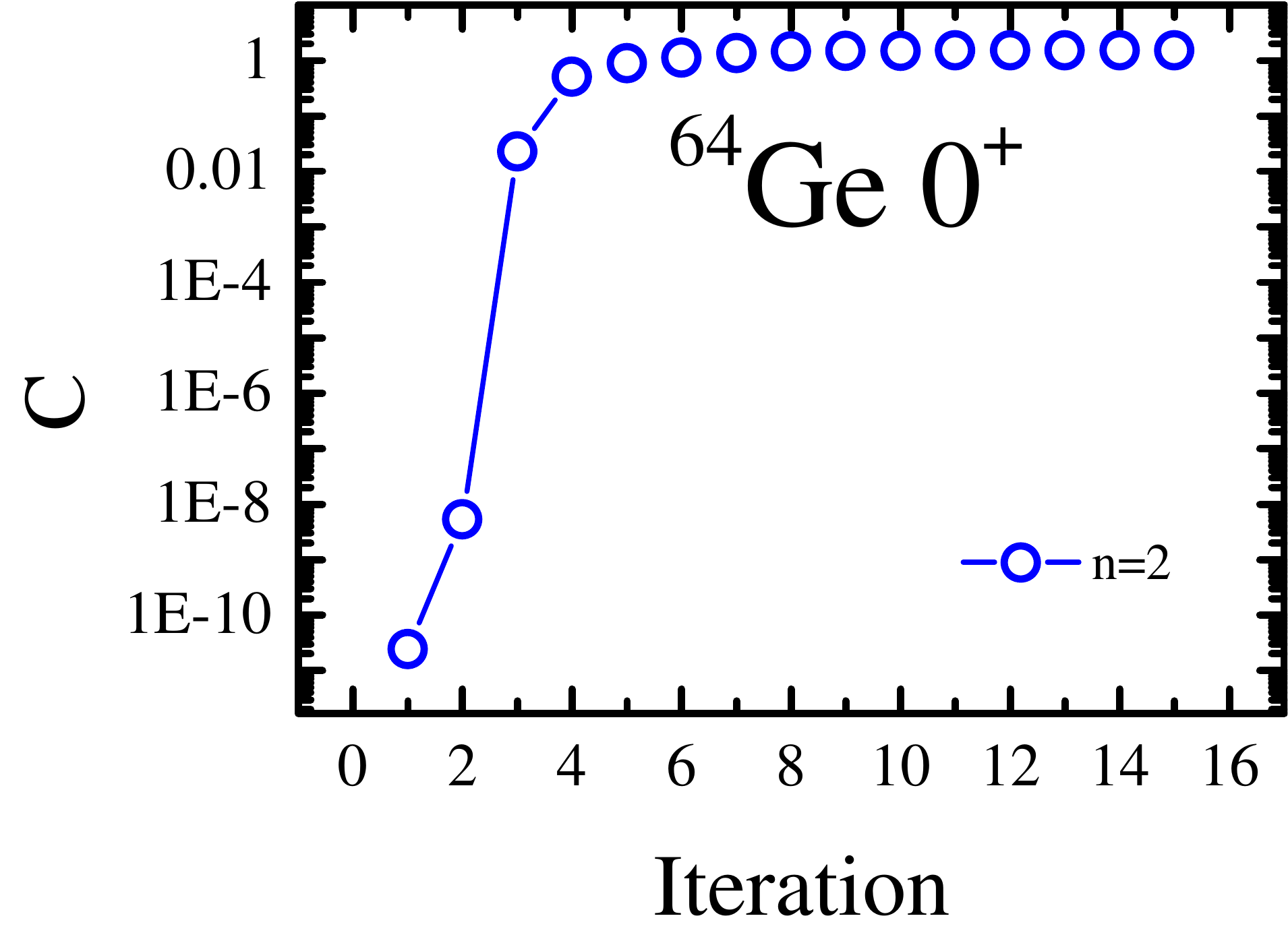}
 \caption{\label{C}(Color online) Iteration of the $C$ quantity in Eq. (\ref{cg}) between the second candidate projected state and the $n=1$ converged VAP wave function in Fig. \ref{Ge64} for the $J^\pi=0^+$ ground state in $^{64}$Ge.
}
\end{figure}

The minimization iteration of $-C$ is performed for the second projected SD in the above example. The results are shown in Fig. \ref{C}. It is interesting that although the $C$ value is extremely tiny at the beginning, it becomes quickly large enough that the second projected SD can sufficiently mix with the original VAP wave function with $n=1$. In this sense, we do not need to wait for the converged $C$. According to Fig. \ref{C}, one can simply take the projected SD at the 4-th iteration so that the computational time for getting the useful projected SD can be considerably saved.

With this useful projected SD $P^{J\pi}_{MK}|\Phi\rangle$, one can then perform the VAP calculation with Eq. (\ref{wf2}). The calculated results are also shown in Fig. \ref{Ge64}. This time, the VAP energy is indeed lowered from $-302.983$MeV to $-303.522$MeV.

With this new method, important projected SDs can be selected one by one so that the VAP iteration can be carried out normally. Thus the VAP wave functions can be continuously improved. At this point, one can naturally imagine that if the VAP process is omitted, then the formed nuclear wave function is very similar to that in the Mont Carlo shell model (MCSM) \cite{MC01}. However, the major difference is, in the MCSM, the important projected SDs are selected from a large amount of the candidates which are generated stochastically. Therefore, it seems that we have proposed an alternative method of basis selection, which can be used to take place of that in the MCSM.

However, before we put more projected states into the VAP wave functions, we should solve another problem that is associated with the orthonormality among the included projected basis states, as will be addressed in the next section.

\section{Orthonormality among the projected basis states}\label{ON}

A natural deficiency of the projected states is the poor orthonormality among themselves. A direct consequence of such deficiency is the possible appearance of the redundant projected states, which seriously affects the stability of VAP iteration. This problem of orthonormality has been addressed in Ref. \cite{VAP22}. In that work, the VAP calculation is performed so that the following $Q$ quantity can be minimized,
\begin{equation}\label{q}
Q=\sum_{\alpha=1}^{m} E_{\alpha}^{J\pi}+\chi_{1} \sum_{i=1}^{n} \frac{1}{N_{i i}^{J\pi}}+\frac{\chi_{2}}{2} \sum_{\substack{i, j=1 \\ i \neq j}}^{n} \frac{N_{i j}^{J\pi} N_{j i}^{J\pi}}{N_{i i}^{J\pi} N_{j j}^{J\pi}},
\end{equation}
where $N_{i j}^{J\pi}=\langle \Phi_i|P^{J\pi}_{KK}|\Phi_j\rangle$. The first term is the sum of the calculated state energies, the last two terms are the constraints which are expected to keep the projected basis states in a good condition so that redundant states may not appear. Actually, one can easily understand that, in Eq. (\ref{q}), the second term tends to push the $N_{i i}^{J\pi}$ norms to large values, and the third term tends to guide the projected basis states to be orthogonal to one another.
Here we propose a more reasonable constraint term that can be used to take place of those two constraint terms in Eq. (\ref{q}) and the new definition of the $Q$ quantity can be written as,
\begin{equation}\label{rd}
Q=\sum_{\alpha=1}^{m} E_{\alpha}^{J\pi}+\chi\frac{1}{|N^{J\pi}|},
\end{equation}
where ${|N^{J\pi}|}$ is the determinant of the norm matrix in Eq. (\ref{hw}).
Notice that this $|N^{J\pi}|$ value equals to the product of all eigenvalues of $N^{J\pi}$. If there is a redundant basis state, then an eigenvalue of $N^{J\pi}$ must be zero. So the $|N^{J\pi}|$ value also must be zero and the constraint term in Eq. (\ref{rd}) becomes infinite which is impossible. This means the redundant state can be strictly forbidden by the new constraint term in Eq. (\ref{rd}). In contrast, the original constraints in Eq. (\ref{q}) does not have such strong power in forbidding the appearance of redundant states. For instance, suppose there are three normalized basis states $|\phi_1\rangle$, $|\phi_2\rangle$ and $|\phi_3\rangle$, with the following relation
$\langle\phi_1|\phi_2\rangle=0$ and $|\phi_3\rangle=\frac{1}{\sqrt{2}}(|\phi_1\rangle+|\phi_2\rangle)$. Then $|\phi_3\rangle$ is clearly a redundant state. In this case,
one can easily get the value of the last term in Eq. (\ref{q}), and it turns out to be $\sqrt{2}\chi_2$, which seems not large enough to strongly restrict the basis states. Furthermore, such $\frac{1}{|N^{J\pi}|}$ term tends to increase of the eigenvalues of $N^{J\pi}$, which also plays the role similar to the second term in Eq. (\ref{q}). Therefore, it is very nice to take the new constraint to keep the precision of the calculated VAP wave functions.

In the present VAP calculations, the $Q$ value in Eq. (\ref{rd}) needs to be minimized. Thus it would be better if the gradient and the Hessian matrix of the $\frac{1}{|N^{J\pi}|}$ term are calculated. Formulations of how to evaluate such quantities have been explicitly presented in the Appendix attached below. Fortunately, all the required matrix elements for the $\frac{1}{|N^{J\pi}|}$ term are actually available because they are originally prepared for the gradient and the Hessian matrix of the VAP energies.

To show the effect of this new  $\frac{1}{|N^{J\pi}|}$ constraint, we perform the VAP calculations in the $sd$ model space. Since this model space is quite small, the dimension of the corresponding configuration space is not so large that the randomly selected projected states can easily mix with one another. This makes it very convenient that one can avoid the complexity of combining this constraint with the first problem addressed in the above section. Therefore, one can randomly generate a group of the projected SDs and directly perform the VAP calculation with or without the $\frac{1}{|N^{J\pi}|}$ constraint.

\begin{figure}[ht!]
 \centering
 \includegraphics[width=8cm]{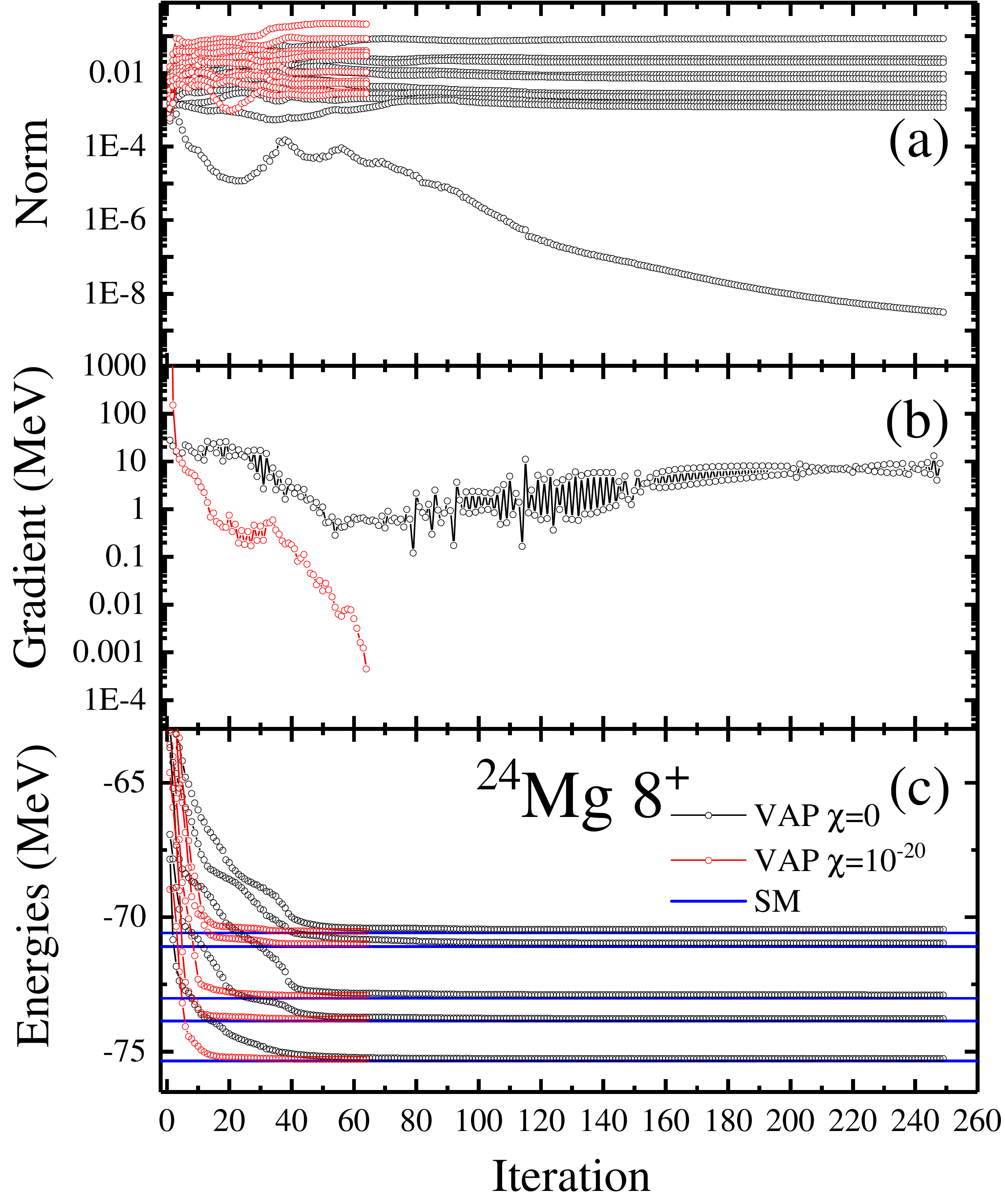}
 \caption{\label{mg248+}(Color online) Calculated quantities as functions of the VAP iteration for the lowest 5 states ($m=5$) with $J^\pi=8^+$ in $^{24}$Mg. 10 projected SDs are adopted to construct the VAP wave functions. The USDB interaction is adopted. (a) The calculated 10 eigenvalues of $N^{J\pi}$; (b) The absolute value of the gradient of the $Q$ quantity;  and (c) the calculated lowest five energies. The shell model (SM) energies are also shown for comparison.
}
\end{figure}

We randomly generate 10 projected SDs ($n=10$) with $K=0$ to construct the lowest 5 $J^\pi=8^+$ states ($m=5$) in $^{24}$Mg. The USDB interaction \cite{usdb} is adopted. Then the $Q$ quantity in Eq. (\ref{rd}) is minimized with $\chi=0$ and $\chi=0.01^{n}=10^{-20}$, respectively. The associated quantities as functions of the VAP iteration are shown in Fig. \ref{mg248+}. It is seen that without any constraint, the lowest eigenvalue of the $N^{J\pi}$ shown in Fig. \ref{mg248+}(a) tends to be smaller and smaller which is very bad in obtaining precise VAP wave functions. Therefore, the gradient of $Q$ in Fig. \ref{mg248+}(b) is not accurately calculated and is very difficult to be smaller than $10^{-3}$MeV, which is the condition of our VAP convergence. Now, let's perform the same calculation but with $\chi=10^{-20}$. This time the eigenvalues of the $N^{J\pi}$ becomes not so small and the VAP iteration converges quite fast. The VAP energies are compared with the exact shell model ones and good approximation still can be achieved even with the new constraint.

\section{The example of the 4$^+$ states in $^{56}$Ni}\label{56ni}
 Generally, the above two problems may appear simultaneously in a VAP calculation. Suppose that there are $m$ lowest states that need to be calculated. The VAP calculation can be performed in two steps. The first step is one can generate $n=m$ projected SDs randomly and construct the simplest VAP wave functions for these $m$ lowest states. Then these wave functions are varied simultaneously so that the $Q$ quantity with $\chi\neq 0$ in Eq. (\ref{rd}) can be minimized . The second step is to improve the VAP wave functions by adding an useful projected state using the algorithm addressed in Section \ref{BS}. Then one can perform the same VAP iteration as that in the first step but with $n=m+1$. In this way, more projected SDs can be effectively added one by one and the approximations of the calculated $m$ lowest states can be continuously improved.

\begin{figure}[ht!]
 \centering
 \includegraphics[width=8cm]{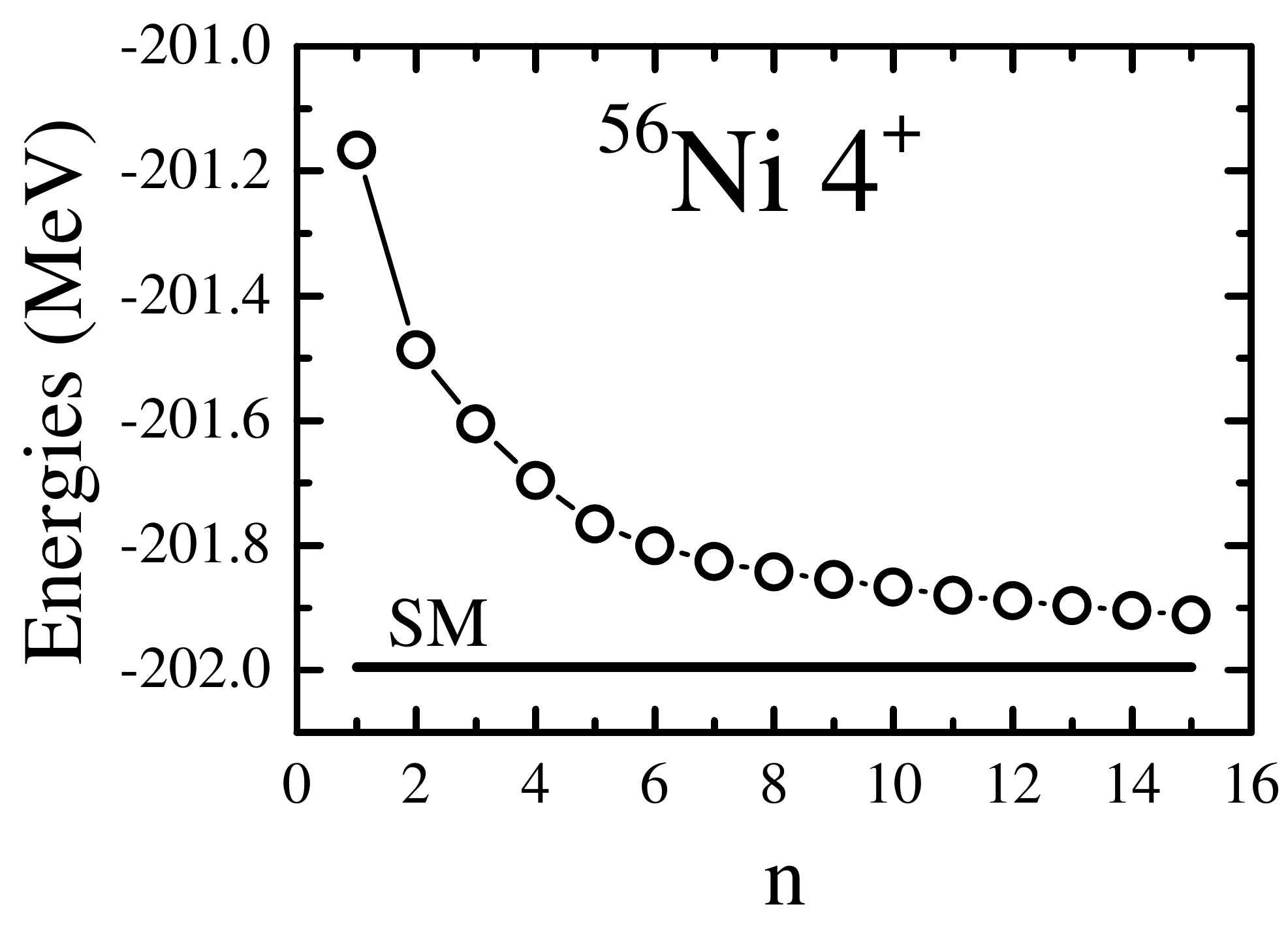}
 \caption{\label{nsd1}(Color online) The lowest (yrast) energy of $J^\pi=4^+$ in $^{56}$Ni calculated with the present VAP as a function of $n$, the number of included projected basis states. The shell model (SM) energy is shown for comparison.
}
\end{figure}

As a practical example, we calculate the lowest $J^\pi=4^+$ states in $^{56}$Ni. The GXPF1A interaction \cite{gxpf1a} in the $pf$ shell model space is adopted and $K=0$ is taken. The constraint parameter $\chi$ is assumed to be associated with the number of the included projected states and is simply taken to be $\chi=10^{-2n}$. Let's first calculate the yrast state ($m=1$). The results are shown in Fig. \ref{nsd1}. It is shown that the calculated energy is lowered steadily with the projected SDs added one by one. This clearly indicates each of the added projected states is indeed important for the lowest $J^\pi=4^+$ state. One can also see that, the energy decrease rapidly at the beginning and then becomes slowly and slowly as the $n$ number increases. This can be understood that, at each $n$, the most important projected state should be obtained because it lowers the energy to the maximum in the VAP algorithm.  This means, the next added projected state should be less important than the previous one.

\begin{figure}[ht!]
 \centering
 \includegraphics[width=8cm]{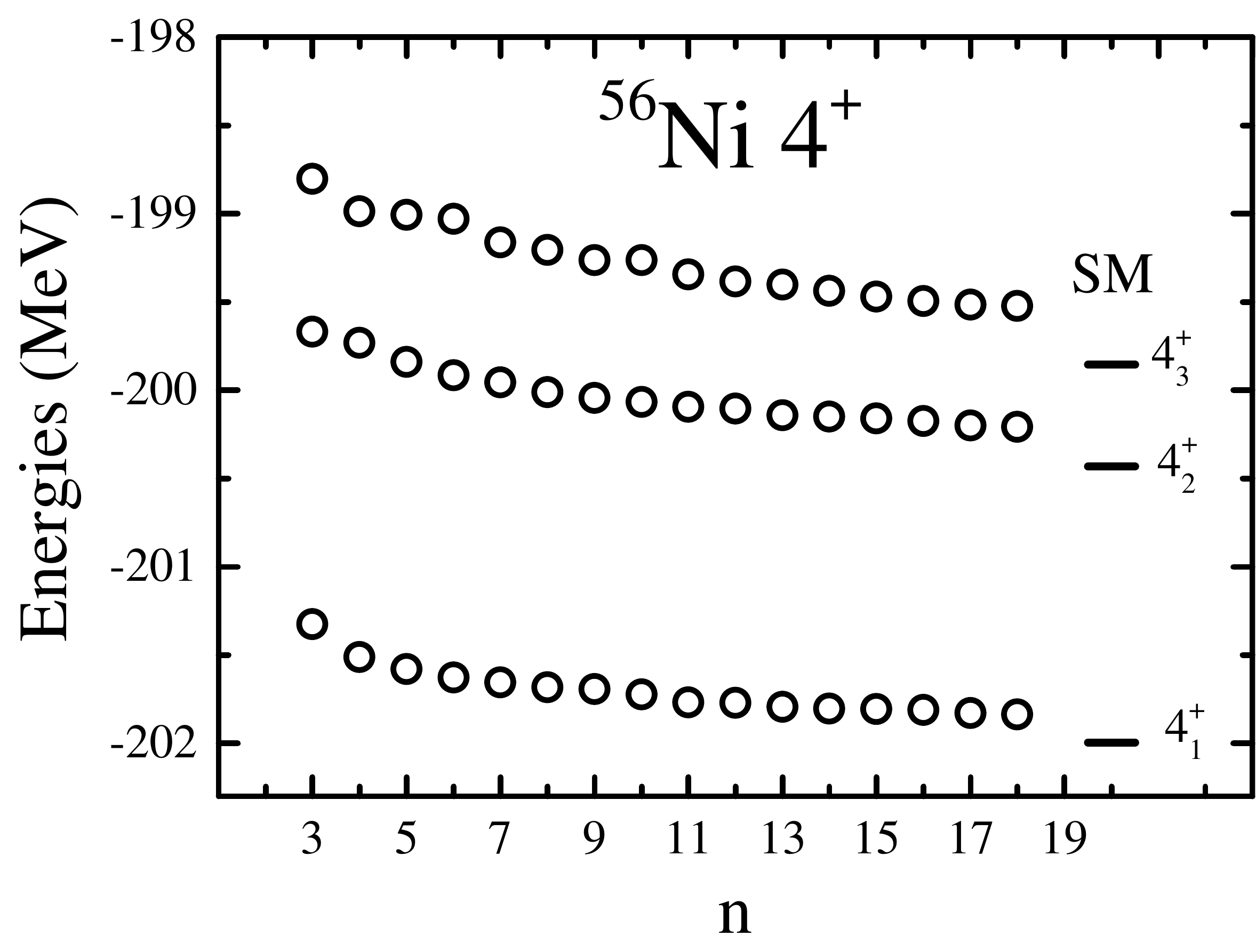}
 \caption{\label{nsd3}(Color online) Similar to the Fig. \ref{nsd1} but for the lowest three energies of $J^\pi=4^+$ in $^{56}$Ni.
}
\end{figure}

Next, let's calculate the lowest three $4^+$ states in $^{56}$Ni, simultaneously.  The calculated results are shown in Fig. \ref{nsd3}. It is seen that all the calculated energies becomes lower and lower as $n$ increases, which seems very similar to Fig. \ref{nsd1}. However, one can also see in Fig. \ref{nsd3}, the energies are not always significantly improved by adding a new projected state. As we have addressed above, if the added new projected state does not mix with one of the state wave functions, then the corresponding energy remains unchanged. From Fig. \ref{nsd3}, one can understand that the selected new projected state can be important at least for one of the calculated states, but may not always be important for all of them. Nevertheless, It is believed the calculated three energies may be sufficiently close to the exact shell model ones if $n$ is larger enough.

\section{Summary and outlook}\label{sum}

The projected wave functions with good quantum numbers are very nice blocks for the construction of the nuclear wave functions, which can be very close to the exact shell model ones. The full optimized nuclear wave functions expanded in terms of these projected states can be obtained through the VAP calculation. However, when the VAP calculations are performed in a large model space, the selected projected basis states are likely scattered in the extremely huge configuration space. This implies that the selected projected state might be useless for the calculated states. To solve this problem, we propose an algorithm that the useless projected states can be varied so that it can be important for the calculated states. Another problem is, the projected basis states are far from the orthonormality, which seriously damages the stability of the VAP iteration. Here, we solve this problem by attaching a new constraint to the sum of the calculated energies. The proposed solutions for the discussed problems have been supported by the calculated examples.

 Certainly, the discussed problems are general ones and are independent of the specific form of the projected states. If the VAP wave functions are formed with the projected Hartree-Fock-Bogoliubov vacuum states, one may still confront with these two problems. In that case, we expect the present solutions would be still valid, which will be further investigated in the future.

\begin{acknowledgments}
This work is supported by the National Natural Science Foundation of China under Grant No. $11975314$, by the Key Laboratory of Nuclear Data foundation (JCKY2022201C158) and by the Continuous Basic Scientific Research Project Nos. WDJC-$2019$-$13$, BJ$20002501$.
\end{acknowledgments}

\appendix
\section{The Gradient and Hessian matrix of the constrain term}
For simplicity, let's rewrite the constraint term as follows,
\begin{equation}\label{Q0}
Q_0=\frac{1}{\left|\begin{array}{lll}
N_{11} &\cdots& N_{1n}\\
 &\cdots& \\
N_{n1} &\cdots& N_{nn}
\end{array}\right|}=\frac{1}{|N| },
\end{equation}
where $N_{i j}=\langle \Phi_i|P^{J\pi}_{KK}|\Phi_j\rangle$. In the VAP method, one can vary the $|\Phi_i\rangle$ states by applying the Thouless theorem to find the best set of VAP basis states, namely  \cite{Ring80}
\begin{equation}\label{thls}
|\Phi_i\rangle=\mathcal{N} e^{\frac{1}{2} \sum_{\mu \nu} d_{\mu \nu} \beta_{0,\mu }^{\dagger}\beta_{0,\nu }^{\dagger}}|\Phi_{0}\rangle,
\end{equation}
where $\mathcal{N}$ is the normalization parameter. $\beta_{0,\mu }^{\dagger}\beta_{0,\nu }^{\dagger}$ is generally a quasiparticle pair operator corresponding to the fixed $|\Phi_{0}\rangle$ HFB vacuum state, but here it refers to a particle-hole operator and $|\Phi_{0}\rangle$ is a Slater determinant. $d$ is a complex skew matrix. The matrix elements $d_{\mu \nu}$ can be a complex number,
\begin{equation}
d_{\mu \nu}=x_{\mu \nu}+i y_{\mu \nu}.
\end{equation}
where, $x_{\mu \nu}$ and $y_{\mu \nu}$ are real numbers and are the variational parameters.
For convenience, we use $x_\alpha$, $x_\beta$, etc. to stand for these variational parameters.

Then, the gradient of $Q_0$ can be expressed as
\begin{eqnarray}\label{GQ0}
\frac{\partial Q_0}{\partial x_{\alpha}}&=&-\frac{1}{|N |^2}\frac{\partial   |N | }{\partial x_{\alpha}}\nonumber\\
&=&-\frac{1}{|N |^2}\sum_{i=1}^n\left|\begin{array}{lll}
N_{11} &\cdots& N_{1n} \\
 &\cdots& \\
\frac{\partial N_{i1} }{\partial x_{\alpha}} &\cdots& \frac{\partial N_{in} }{\partial x_{\alpha}} \\
  &\cdots& \\
N_{n1}  &\cdots& N_{nn}
\end{array}\right|\nonumber\\
&=&-\frac{1}{|N |^2}\sum_{i,k=1}^n\frac{\partial N_{i k}  }{\partial x_{\alpha}}\overline{N }\{i|k\} ,
\end{eqnarray}
where $\overline{N }\{i|k\}=(-1)^{i+j}|N \{i|k\}|$ is a cofactor of $|N |$. The submatrix $N \{i|k\}$ is obtained by removing the $i$th row and $k$th column from the matrix $N $. Calculation of $\frac{\partial N_{i k}  }{\partial x_{\alpha}}$ can be found in Ref. \cite{VAP17}.

Similarly, the Hessian of $Q_0$ can be written as
\begin{eqnarray}\label{HQ0}
\frac{\partial^2 Q_0}{\partial x_{\alpha}\partial x_{\beta} }&=&2\frac{1}{|N |^3}\frac{\partial |N | }{\partial x_{\alpha}}\frac{\partial |N |}{\partial x_{\beta}}-\frac{1}{|N |^2}\frac{\partial^2 |N | }{\partial x_{\alpha}\partial x_{\beta}},
\end{eqnarray}
where the $\frac{\partial^2 |N | }{\partial x_{\alpha}\partial x_{\beta}}$ in the second term of Eq. (\ref{HQ0}) can be calculated with the following equation,
\begin{eqnarray}
\frac{\partial^2 |N | }{\partial x_{\alpha}\partial x_{\beta}}
&=&\sum_{i=1}^n\left|\begin{array}{lll}
N_{11}  &\cdots& N_{1n} \\
 &\cdots& \\
\frac{\partial^2 N_{i1} }{\partial x_{\alpha}\partial x_{\beta}} &\cdots& \frac{\partial ^2N_{in} }{\partial x_{\alpha}\partial x_{\beta}} \\
  &\cdots& \\
N_{n1}  &\cdots& N_{nn}
\end{array}\right| \nonumber\\
&&+\sum_{i,j=1\atop i\neq j}^n\left|\begin{array}{lll}
N_{11}  &\cdots& N_{1n} \\
 &\cdots& \\
\frac{\partial  N_{i1} }{\partial x_{\alpha} } &\cdots& \frac{\partial N_{in} }{\partial x_{\alpha} } \\
  &\cdots& \\
\frac{\partial N_{j1} }{ \partial x_{\beta}} &\cdots& \frac{\partial N_{in} }{ \partial x_{\beta}} \\
  &\cdots& \\
N_{n1}  &\cdots& N_{nn}
\end{array}\right| \nonumber\\
&=&\sum_{i,k=1}^n\frac{\partial^2 N_{i k}  }{\partial x_{\alpha}\partial x_{\beta}}\overline{N }\{i|k\} \nonumber\\
&&+ \sum_{ijkl,\atop i\neq j, (k<l)}\left|\begin{array}{ll}
\frac{\partial N_{ik} }{\partial x_{\alpha}} &  \frac{\partial N_{il} }{\partial x_{\alpha}}\\
\frac{\partial N_{jk} }{\partial x_{\beta}} &  \frac{\partial N_{jl} }{\partial x_{\beta}}
\end{array}\right|\overline{N }\{ij|kl\}. \nonumber\\
\end{eqnarray}
The $\frac{\partial^2 N_{i k}  }{\partial x_{\alpha}\partial x_{\beta}}$ also can be found in Ref. \cite{VAP17}.
If $i < j$ and $k < l$, then $\overline{N }\{ij|kl\}$ is usually called as the second order cofactor define by
\begin{eqnarray}
\overline{N }\{ij|kl\}=(-1)^{i+j+k+l}|N \{ij|kl\}|,
\end{eqnarray}
where the submatrix $N \{ij|kl\}$ is obtained from the $N $ matrix by removing the $i,j$th rows and $k,l$th columns.

Both $\overline{N }\{i|k\}$ and $\overline{N }\{ij|kl\}$ can be easily obtained by using Jacobi's identity in the matrix theory \cite{jacobi},
\begin{eqnarray}
\overline{N }\{i|k\}={N_{ki} }^{-1}|N  |,
\end{eqnarray}
\begin{eqnarray}
\overline{N }\{ij|kl\}= \left|\begin{array}{ll}
  {N_{ki} }^{-1} &  {N_{kj} }^{-1}\\
{N_{li} }^{-1} &  {N_{lj} }^{-1}
\end{array}\right||N |.
\end{eqnarray}

{}
 
\end{document}